\documentclass[prl,twocolumn,amsmath,amssymb,bibnotes,aps,longbibliography]{revtex4-1}

\usepackage{textcomp}
\usepackage{pifont}
\usepackage{mathptmx}
\usepackage{graphicx}
\usepackage{enumerate}
\usepackage{titlesec}
\usepackage{color}
\usepackage[normalem]{ulem}
\usepackage{isotope}
\usepackage{hyperref}
\hypersetup{
pdftitle={Detecting an Itinerant Optical Photon Twice without Destroying It},
pdfauthor={E. Distante, S. Daiss, S. Langenfeld, L. Hartung, P. Thomas, O. Morin, G. Rempe, and S. Welte},
colorlinks=true, linkcolor=[RGB]{45,48,146}, citecolor=[RGB]{45,48,146}, filecolor=[RGB]{45,48,146}, urlcolor=[RGB]{45,48,146}}

\newcommand{\ket}[1]{|{#1}\rangle}
\newcommand{\braket}[1]{\langle{#1}\rangle}
\DeclareTextFontCommand{\emph}{\textit}
\makeatletter 
\@ifpackageloaded{hyperref}{\newcommand{\positionlabel}[1]{\hypertarget{#1}{}}}{\newcommand{\positionlabel}[1]{}}
\@ifpackageloaded{hyperref}{}{\newcommand{\hyperref}[2][]{#2}}
\@ifpackageloaded{hyperref}{}{}
\makeatother

\begin{document}
\title{Detecting an Itinerant Optical Photon Twice without Destroying It}

\author{Emanuele~Distante$^{1}$}
\author{Severin~Daiss$^{1}$}
\author{Stefan~Langenfeld$^{1}$}
\author{Lukas~Hartung$^{1}$}
\author{Philip~Thomas$^{1}$}
\author{Olivier~Morin$^{1}$}
\author{Gerhard~Rempe$^{1}$}
\author{Stephan~Welte$^{1}$}
\email[To whom correspondence should be addressed. Email: ]{stephan.welte@mpq.mpg.de}
\affiliation{$^{1}$Max-Planck-Institut f{\"u}r Quantenoptik, Hans-Kopfermann-Strasse 1, 85748 Garching, Germany}

\begin{abstract}
\noindent Nondestructive quantum measurements are central for quantum physics applications ranging from quantum sensing to quantum computing and quantum communication. Employing the toolbox of cavity quantum electrodynamics, we here concatenate two identical nondestructive photon detectors to repeatedly detect and track a single photon propagating through a $60\,\mathrm{m}$ long optical fiber. By demonstrating that the combined signal-to-noise ratio of the two detectors surpasses each single one by about two orders of magnitude, we experimentally verify a key practical benefit of cascaded non-demolition detectors compared to conventional absorbing devices.
\end{abstract}

\maketitle \nocite{apsrev41Control}

\noindent Quantum physics distinguishes between two kinds of measurements. Following Pauli \cite{pauli1933}, a measurement of the first kind projects the state of a system onto an eigenstate of the measured observable, with subsequent measurements of the same kind giving the same result. A measurement of the second kind, in contrast, exerts a random back-action on the complementary observable so that repeated measurements lead to different results. This distinction has been refined by introducing the concept of a quantum non-demolition (QND) measurement \cite{braginsky1980,braginsky1996}, where Pauli's measurement of the first kind is formally defined by requesting that the operator corresponding to the measured observable commutes with the Hamiltonian of the system. In practice, QND measurements thus allow for repeated observations without changing the outcome, a unique property that has (at least) two benefits. First, it allows to track the evolution of the observable without any back-action. Second, it allows to concatenate several measurements, each with a non-perfect detection sensitivity (e.g., signal-to-noise ratio), and in this way enhance the overall sensitivity. Repeatability therefore has been identified a powerful advantage, as was emphasized by Caves \textit{et al.} \cite{caves1980}: ``The key feature of such a non-demolition measurement is \textit{repeatability} -- once is not enough!".

The application potential of QND measurements was realized early on in the field of  gravitational wave detection \cite{braginsky1980,braginsky1996} and has later on sparked large interest in a vast number of other fields ranging from astronomy \cite{kellerer2014quantum}, to high-precision metrology \cite{giovanetti2004, ma2011} and quantum-information processing \cite{ralph2006}. In the laboratory, QND measurements have been implemented in different matter-based experimental platforms such as ions \cite{hume2007}, superconducting qubits \cite{lupacscu2007}, solid state systems \cite{neumann2010}, and atomic ensembles \cite{kuzmich2000}. QND measurements of single photons, however, turned out to be comparatively difficult to implement, as they require the development of detectors capable of observing single photons without absorbing them \cite{braginsky1989}. Nevertheless, landmark experiments were proposed and performed in the microwave domain with photons stored in high-quality resonators \cite{brune1992, nogues1999seeing, guerlin2007} and later extended to the detection of itinerant microwave photons \cite{kono2018quantum, besse2018}. In the optical domain \cite{friberg1992, grangier1998}, a single nondestructive detection of a flying photon was achieved as well \cite{reiserer2013}. Although this experiment demonstrated the principle of one QND detection, so far an experimental verification of the feasibility to repeatedly measure and thereby track a flying optical photon remained elusive. 

\begin{figure}[b]
\centering
\includegraphics[width=\columnwidth]{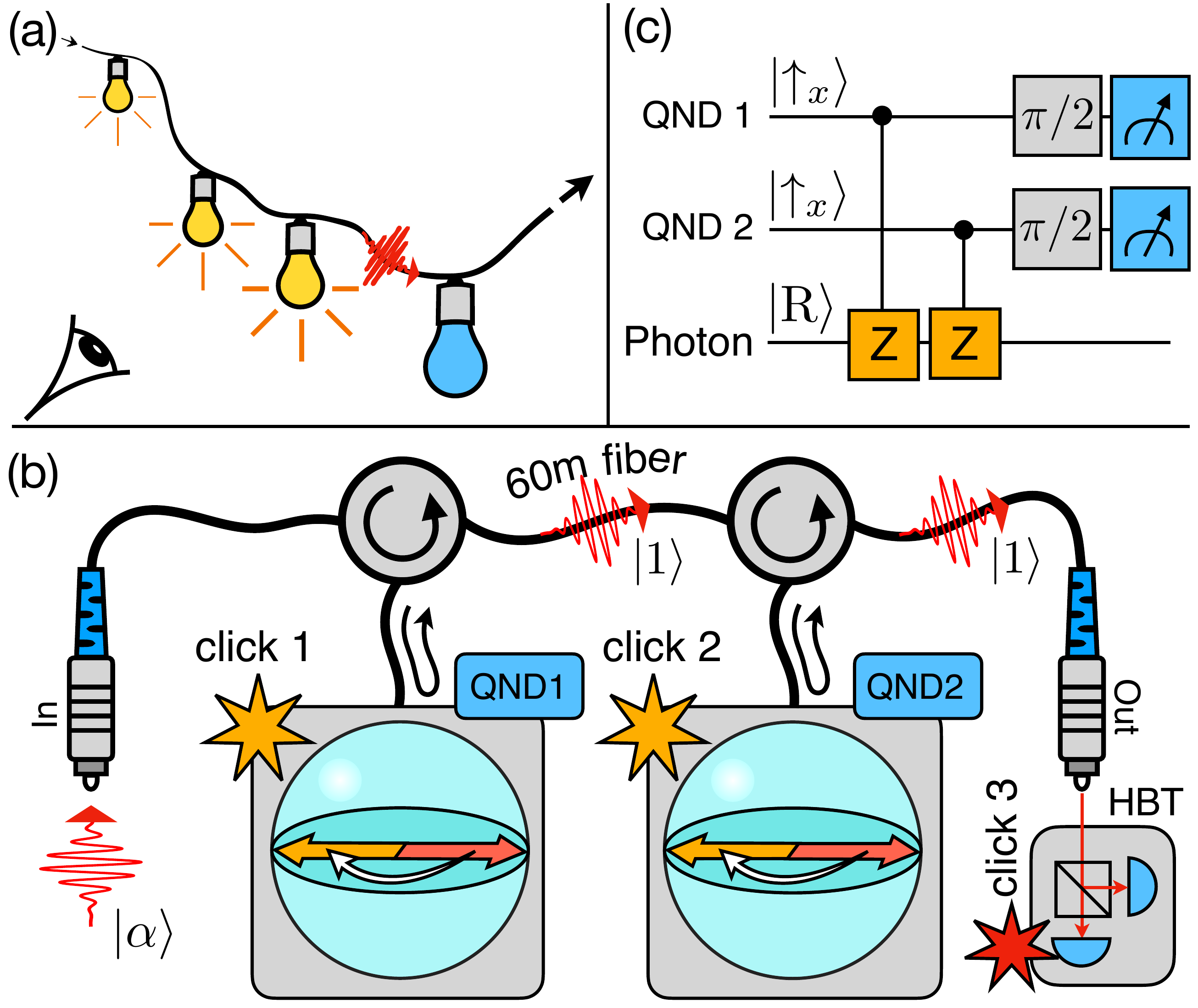}
\caption{\label{fig:setup} (a) Our vision. A series of QND detectors, depicted as bulbs, are attached to an optical fiber. A propagating photon (red wiggly arrow) is subsequently detected several times, indicated by the illuminated bulbs. (b) Setup of the experiment. A weak coherent pulse $\ket{\alpha}$ is coupled into an optical fiber and consecutively interacts with the two QND detectors. The latter are comprised of an atomic qubit each, depicted as a Bloch sphere (light blue). The north/south poles of the Bloch spheres correspond to $\ket{\uparrow_z}$ and $\ket{\downarrow_z}$, respectively. The states $\ket{\uparrow_x}$ $(\ket{\downarrow_x})$ on the equator are indicated by the red (yellow) arrows. Two circulators direct the photons onto the QND detectors before they are directed to a Hanbury Brown-Twiss (HBT) setup of single-photon detectors (blue half disks). (c) Quantum circuit diagram. The photon reflections from the two detectors comprise two Z-gates.}
\end{figure}

Here we verify the repeatability and the increased sensitivity by using two QND detectors to observe one optical photon twice. The two devices, each made from a single atom coupled to an optical cavity, are distributed along an optical fiber in which the photon propagates. Once the latter has interacted with each detector, we observe correlations between the detection events. These correlations are the key element behind our demonstration that the combined system of two detectors outperforms both individual devices in terms of signal-to-noise ratio. Furthermore, we demonstrate that our QND detectors are specifically suited to detect single-photon states. To this end, we employ the first QND detector as a state-preparation device to generate single photons out of weak coherent states \cite{daiss2019}. We then use such single photons to probe the second QND detector. As they are eigenstates of the measurement, we show that they are detected with higher probability than the input coherent states and that their single-photon character is unaffected by the QND detection. Our protocol is widely applicable and could also be implemented with single ions \cite{mundt2002, takahashi2020}, superconducting qubits \cite{kono2018quantum, besse2018}, quantum dots \cite{fushman2008, kim2013, desantis2017}, silicon vacancy centers \cite{bhaskar2020} or rare-earth ions \cite{chen2020} coupled to cavities.

The setup consists of a concatenation of QND detectors, lined up along an optical fiber in which a photon propagates. An artist's view of the experiment is shown in Fig. \ref{fig:setup}(a) where the QND detectors are depicted as bulbs that light up as soon as the photon passes by. An external observer can see correlations between the successively illuminated bulbs, and photon loss manifests by dark bulbs downstream. Fig. \ref{fig:setup}(b) shows a more detailed sketch of our experiment. The two detectors, named QND1 and QND2 in the following, are separated by a distance of $60\,\mathrm{m}$. Each detector is made of a single \isotope[87]{Rb} atom trapped in a high-finesse cavity. The systems are both single-sided and operate in the strong-coupling regime of cavity quantum electrodynamics (CQED). The atoms are initialized in $\ket{\uparrow_z}=\ket{5\isotope[2]{S}_{1/2}, F=2, m_F=2}$ via optical pumping with light resonant with the $\ket{\uparrow_z}\leftrightarrow \ket{e}$ transition, where $\ket{e}=\ket{5\isotope[2]{P}_{3/2}, F=3, m_F=3}$. The cavities are both stabilized to this transition frequency. Following the optical pumping, we employ a pair of Raman lasers in each of the setups to prepare both atoms in the
superposition state $\ket{\uparrow_x}=\frac{1}{\sqrt{2}}(\ket{\uparrow_z}+\ket{\downarrow_z})$, where $\ket{\downarrow_z}=\ket{5\isotope[2]{S}_{1/2}, F=1, m_F=1}$ \cite{reiserer2013}. A state-detection laser resonant with the $\ket{\uparrow_z}\leftrightarrow\ket{e}$ transition allows to deterministically distinguish between the states $\ket{\uparrow_z}$ and $\ket{\downarrow_z}$ with a fidelity of $>99\%$.

Instead of injecting single photons into the fiber, we perform our experiments with weak coherent laser pulses $\ket{\alpha}$ that contain a mean photon number $\vert\alpha\vert^2=\braket{n}$ in front of the first QND detector. The choice of coherent pulses ($\lambda=780\,\mathrm{nm}$) allows us to study the application of our QND detectors as quantum state preparation devices of single photons. Additionally, in the limit of $\braket{n}\rightarrow 0$, we can approximate a single-photon input by eliminating the vacuum contribution through a post selection on a detection event (`click') in a standard absorbing detector at the far end of the propagation line. The light is injected into the fiber and, after the reflection from the QND detectors, hits two absorbing detectors arranged in Hanbury Brown-Twiss configuration that allows us to measure the second-order intensity auto-correlation function $g^{(2)}(\tau)$. The light pulses have a Gaussian shape with a full width at half maximum of $1\,\mathrm{\mu s}$ and are resonant with the transition $\ket{\uparrow_z}\leftrightarrow\ket{e}$.

Our protocol, depicted as a quantum circuit diagram in Fig. \ref{fig:setup}(c), starts by initializing each atom in the state $\ket{\uparrow_x}$. As a next step, light injected into the fiber successively interacts with the two QND detectors. In case of an odd number of photons in the light pulse, as for a single photon, a phase shift of $\pi$ in the combined atom-light state leads to a sign change in both atomic superposition states \cite{duan2004, xiao2004, reiserer2013, tiecke2014}. Each of the QND detectors then occupies the state $\ket{\downarrow_x}=\frac{1}{\sqrt{2}}(\ket{\uparrow_z}-\ket{\downarrow_z})$. For an even number of photons, such as for the vacuum state, the two atoms remain in $\ket{\uparrow_x}$. In the quantum-information language, the interaction of the photon and each of the atoms can be expressed as a controlled-Z gate. After reflection of the light from the two detectors, a $\pi/2$ pulse is applied which maps $\ket{\downarrow_x}$ $(\ket{\uparrow_x})$ onto $\ket{\uparrow_z}$ $(\ket{\downarrow_z})$. Therefore, a final atomic state detection of $\ket{\uparrow_z}$ ($\ket{\downarrow_z}$) heralds the presence of an odd (even) photon number at the corresponding detector.

Since the passing photon is ideally detected by both QND detectors, correlations between both of them must be observable. However in practice, the setup exhibits substantial losses between the two detectors (optical fiber coupling, fiber losses and limited circulator transmission, total transmission: $53\%$). To ensure the presence of light in front of each cavity, we can condition our data on the successful transmission of the photon through the entire system by postselecting on a click of one of our absorbing detectors downstream.

In a first experiment, we individually characterize the two QND detectors by measuring the probability of a successful QND photon detection conditioned on a click in the absorbing detectors. For the measurement, the mean photon number $\braket{n}$ in the impinging coherent pulse is scanned. For $\braket{n}=0.084$, the maximum probabilities $\text{P}(\uparrow_{z, \text{QND1}}\vert\text{click})=81.3\%$ and $\text{P}(\uparrow_{z, \text{QND2}}\vert\text{click})=87.0\%$ are observed. The difference in the measured maximum probabilities $\text{P}(\uparrow_{z, \text{QND1/2}}\vert\text{click})$ is due to slightly different experimental parameters, mainly coherence time and the quality of the Raman pulses, in the two QND detectors. A detailed description of the individual characterization is given in the \hyperref[supplement]{Supplemental Material}.

\begin{figure}[t]
\centering
\includegraphics[width=\columnwidth]{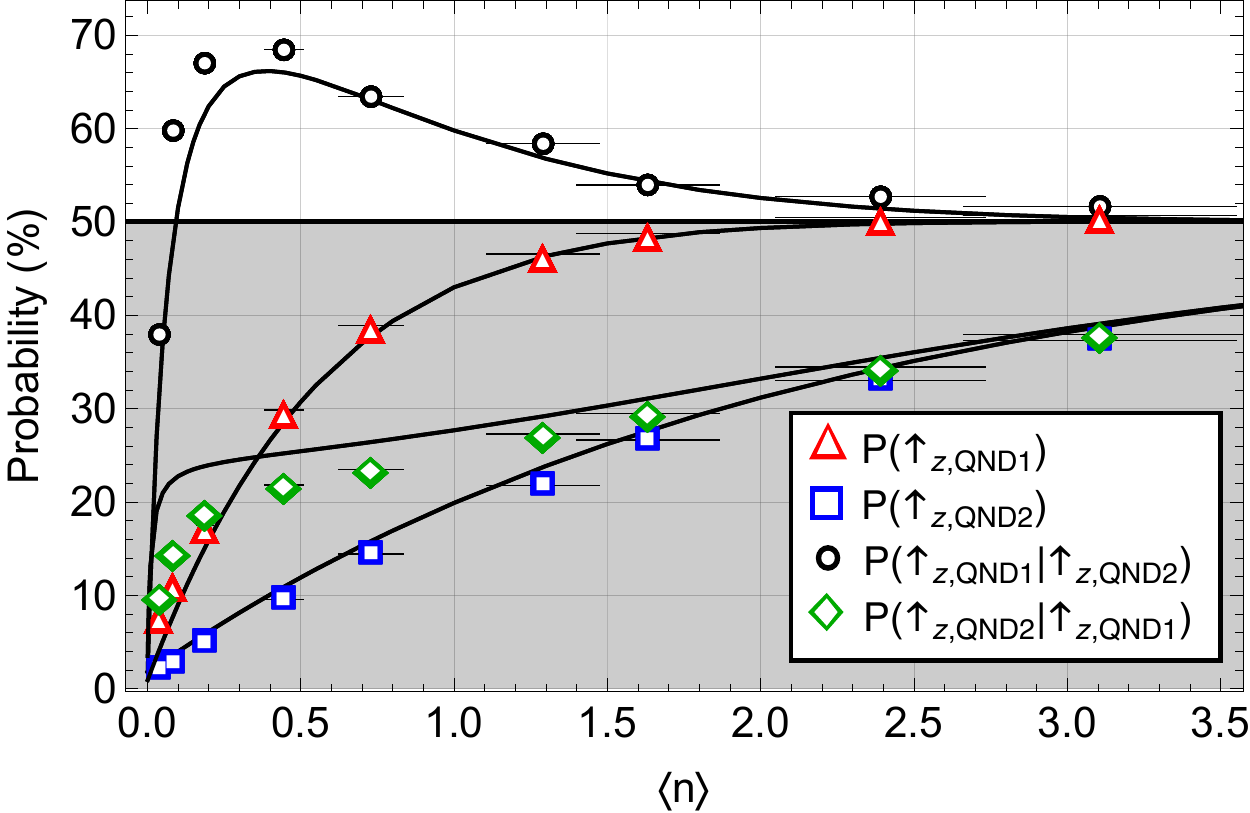}
\caption{\label{fig:correlations} Click probabilities $\text{P}(\uparrow_{z,\text{QND1/2}})$ (red triangles/blue squares) of the two QND detectors when probing them with a coherent state input. When conditioning on a click of the other detector respectively, we observe $\text{P}(\uparrow_{z,\text{QND1/2}}\vert\uparrow_{z,\text{QND2/1}})$ (black circles/green diamonds). The solid horizontal line shows the threshold of $50\%$. The solid black lines are based on a simulation model outlined in the \hyperref[supplement]{Supplemental Material}. The theory is not a fit, but employs experimental parameters characterized in independent measurements. The error bars represent standard deviations from the mean. }
\end{figure}

In the next experiment, we concatenate the QND devices and remove the conditioning on the classical detector click. In a first step, we measure the click probabilities $\text{P}(\uparrow_{z,\text{QND1/2}})$ of the two detectors when probing them with a coherent state containing a mean photon number $\braket{n}$. Since the detectors are sensitive to the parity of the photon number, the respective click probability monotonically increases from zero to $50\%$ as $\braket{n}$ is increased starting from zero. The saturation behavior results from the equal contributions of even and odd photon numbers in the limit of high $\braket{n}$. Due to the optical losses between the two setups, the respective mean photon number is different in front of the two detectors resulting in the observed scaling of the respective curves. Data are shown in Fig. \ref{fig:correlations}.

Since the QND detectors are nondestructive, we can condition each of them on a click in the other detector. The respective click probabilities $\text{P}(\uparrow_{z,\text{QND1/2}}\vert\uparrow_{z,\text{QND2/1}})$ show a different behavior compared to $\text{P}(\uparrow_{z,\text{QND1/2}})$. When conditioning the downstream detector on the upstream detector, an increase in the click probability is observable compared to the case where no conditioning is applied. The effect, however, is strongly suppressed due to the inevitable losses between the setups. In the other case, where the upstream detector is conditioned on the downstream detector, the effect of the losses is suppressed as, for $\braket{n} \ll 1$, a click downstream selects the events when no losses occurred between the setups. For increasing $\braket{n}$, we observe a correlation maximum of $\text{P}(\uparrow_{z, \text{QND1}}\vert \uparrow_{z, \text{QND2}})=(68.4\pm0.7)\%$, clearly surpassing the $50\%$ threshold for uncorrelated random click events. For $\braket{n}$ approaching zero, these correlations decrease due to intrinsic dark counts in the QND detectors. The dark counts of QND1(2) stem from imperfect Raman rotations that leave $\text{dc}_{1(2)}=1.4\%(0.4\%)$ residual population in the state $\ket{\uparrow_z}$ when performing two consecutive $\pi/2$ pulses after initializing the atoms in $\ket{\uparrow_z}$. Due to higher photon-number contributions for increasing $\braket{n}$, the correlations asymptotically approach $50\%$.

\begin{figure}[t]
\centering
\includegraphics[width=\columnwidth]{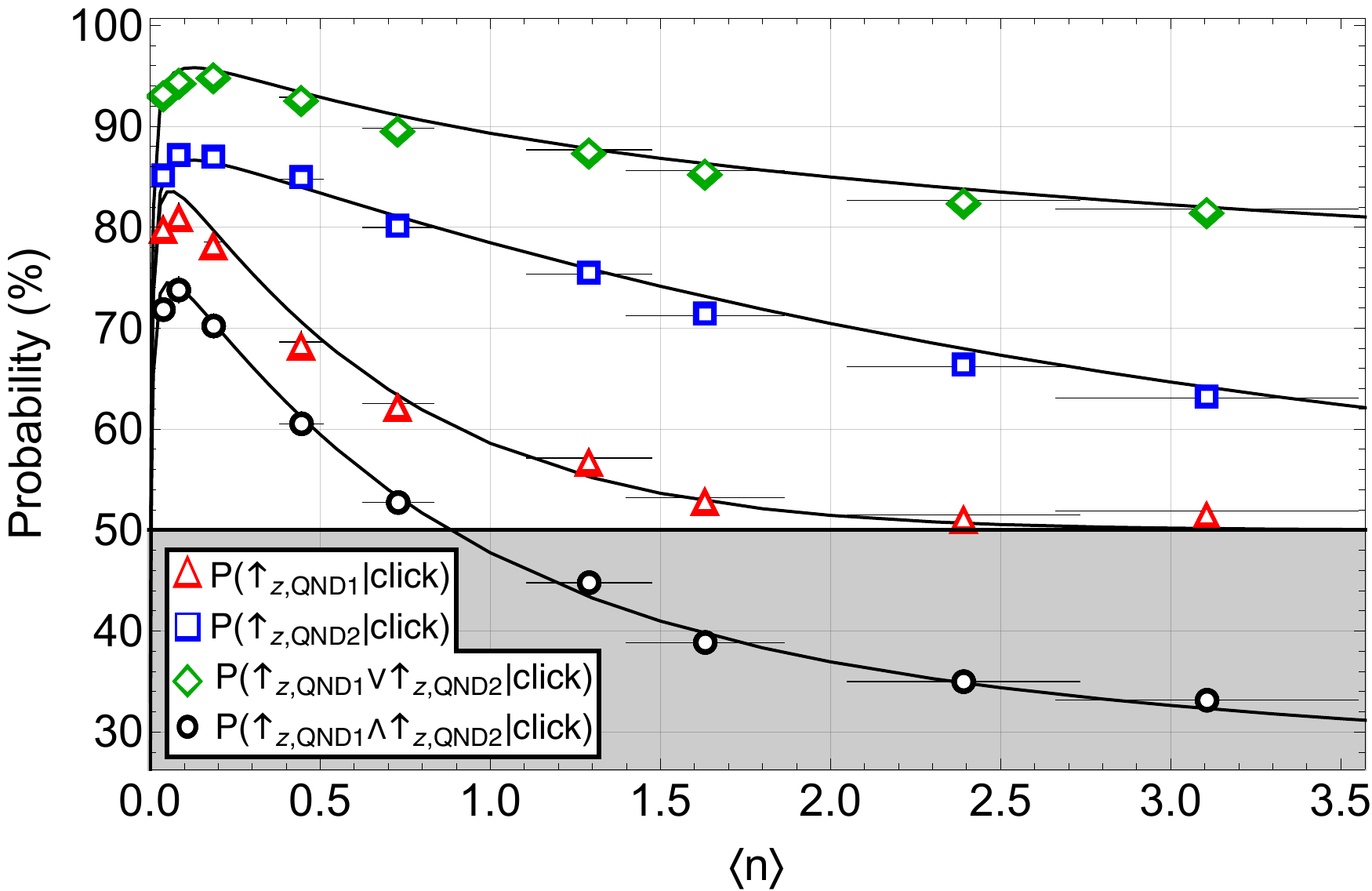}
\caption{\label{fig:efficiency} Correlations between the QND detectors with conditioning on the absorbing-detector click downstream. The green diamonds/black circles, show the probability that QND1 or/and QND2 clicks, conditioned on the absorbing-detector click. The black data points relate to Fig. \ref{fig:setup}(a), where ideally all upstream detectors click once the photon has passed. For comparison, we also show $\text{P}(\uparrow_{z,\text{QND1/2}}\vert\text{click})$ (red triangles/blue squares). Error bars represent standard deviations from the mean. The solid lines show the predictions based on our simulation model.}
\end{figure}

A benefit of nondestructive photon detectors compared to conventional destructive devices is that they can be concatenated to enhance the overall detection efficiency or the signal-to-noise ratio (SNR). In the following, we show that introducing a logical ``or" ($\lor$) connection between the QND detector clicks allows to enhance the overall efficiency while a logical ``and" ($\land$) connection increases the signal-to-noise ratio. We start with the first case and measure the probability that at least one detector, QND1 or QND2, detects a photon conditioned on an absorbing-detector click downstream. A maximum probability of $95.1\%$ is observed which surpasses the maximal capabilities of both individual QND devices ($81.3\%$ and $87.0\%$) and therefore shows the enhancement of the detection efficiency. Data as a function of $\braket{n}$ are shown in Fig.~\ref{fig:efficiency}. A summary of the total measurement time, number of experimental runs, and coincidence rates is given in the \hyperref[supplement]{Supplemental Material}. 

Although the efficiency is an important parameter of a single-photon detector, the signal-to-noise ratio is even more relevant in practice. In the limit of small $\braket{n}$, we define the SNR of the individual detectors as $\text{SNR}_{1(2)}=\text{P}(\uparrow_{\text{z,QND1(2)}}\vert\text{click})/\text{dc}_{1(2)}$ and the SNR of the concatenated detectors as $\text{SNR}_{1\land2}=\text{P}(\uparrow_{\text{z,QND1}}\land\uparrow_{\text{z,QND2}}\vert\text{click})/\text{dc}_{1\land2}$. In the last expression, $\text{dc}_{1\land2}$ is the probability of finding both QND1 and QND2 in $\ket{\uparrow_z}$ at the end of the protocol when no light is injected into the fiber. By employing the logical ``and" connection, we exploit the fact that the signal in both detectors is correlated while the dark counts are uncorrelated to enhance $\text{SNR}_{1\land2}$ compared to $\text{SNR}_{1}=59$ and $\text{SNR}_{2}=218$. While in the limit of small $\braket{n}$ the probability $\text{P}(\uparrow_{\text{z,QND1}}\land\uparrow_{\text{z,QND1}}\vert\text{click})$ is only slightly lowered compared to the individual devices (see Fig.\ref{fig:efficiency}), we find that $\text{SNR}_{1\land2}$ is a factor of 61 higher than $\text{SNR}_2$ and a factor of $227$ higher than $\text{SNR}_1$.

\begin{figure}[t]
\centering
\includegraphics[width=\columnwidth]{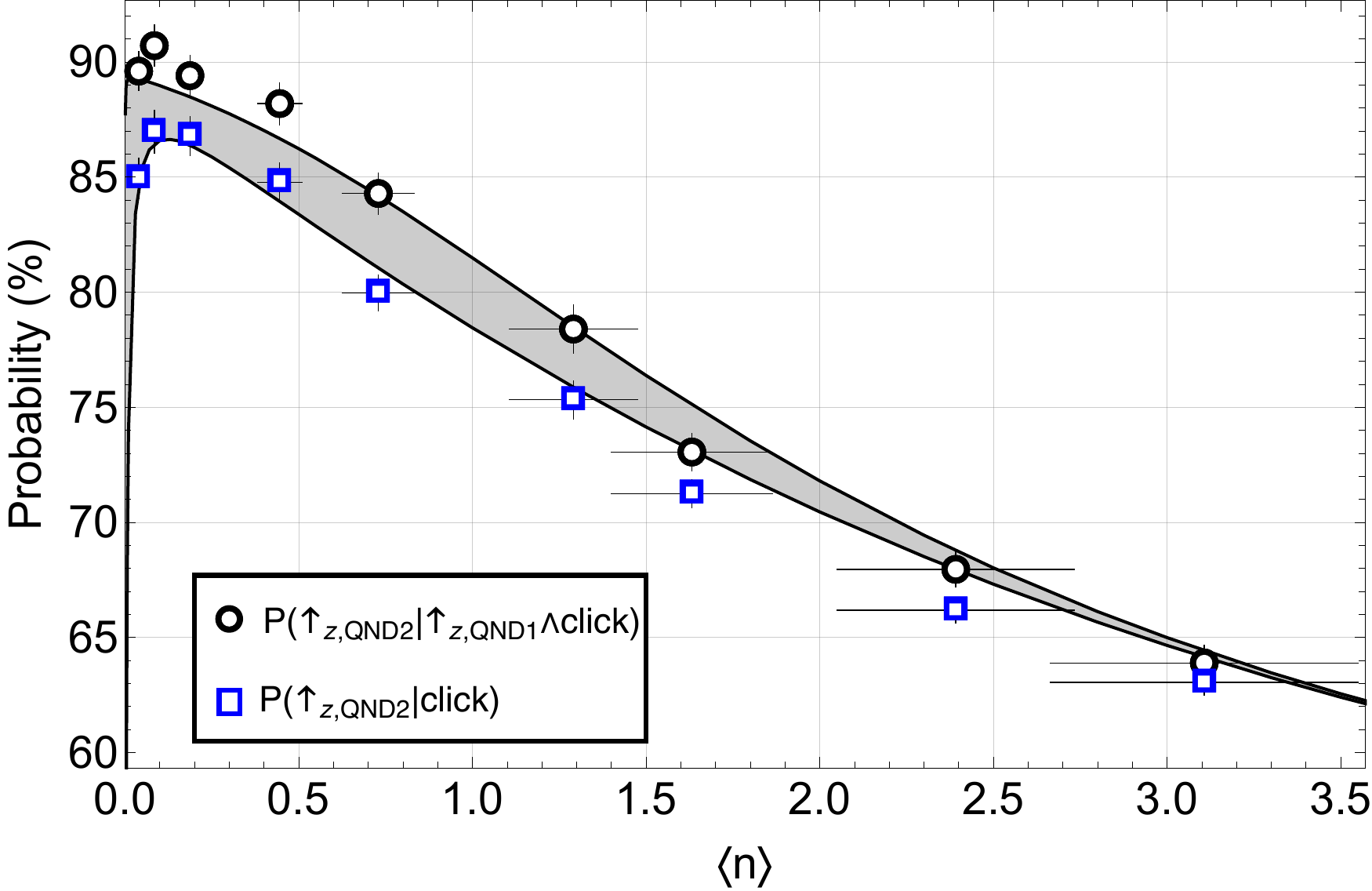}
\caption{\label{fig:distillation}
Probabilities $\text{P}(\uparrow_{z, \text{QND}2}\vert \uparrow_{z,\text{QND}1}\land  \text{click})$ (black circles) and $\text{P}(\uparrow_{z, \text{QND}2}\vert \text{click})$ (blue squares).  The solid lines show theoretical predictions based on our model. The gray area highlights the difference between the two theoretical curves. Error bars represent standard deviations from the mean.
}
\end{figure}

We now demonstrate that increasing the state overlap of the impinging light, $\ket{\alpha}$, with that of a single-photon Fock state, $\ket{1}$, increases the click probability of QND2. This increase stems from the fact that single-photon states are eigenstates of our detector which are observed with unity efficiency in an ideal scenario. Starting with a weak coherent state at the input of the fiber, we employ two conditions ensuring that the light impinging on QND2 is approximately described by a single-photon state. First, the condition on a click in the downstream absorbing detectors is applied. This condition removes the vacuum contribution from the coherent state and additionally ensures that light is not lost in the fiber and therefore impinges on QND2. The limitations of this condition are that all photon-number contributions $n\ge1$ will remain in the pulse, and that for too low values of $\braket{n}$ the dark counts in the absorbing detectors add a small vacuum contribution. Nevertheless, for appropriately chosen low values of $\braket{n}$, this technique allows to approximate single photons. Second, and as an additional condition, we employ QND1 as a state preparation device. Since it is sensitive to the parity of the impinging photon number \cite{wang2005, hacker2019}, a click in this detector removes all even photon-number contributions \cite{daiss2019}. Therefore, in the limit of vanishing $\braket{n}$, employing both conditions applies a selection window around the desired single-photon Fock state and approximates a single photon better than just conditioning on the absorbing detector. As a result, for a proper single-photon detector, the following inequality must hold:  $\text{P}(\uparrow_{\text{z, QND2}}\vert\uparrow_{\text{z, QND1}}\land\text{click})>\text{P}(\uparrow_{\text{z, QND2}}\vert\text{click})$. 
As shown in Fig. \ref{fig:distillation}, we can indeed verify this inequality.
Our data show that the click probability of QND2 increases if the impinging state has a higher overlap with an ideal single-photon state. Maximally, a probability $\text{P}(\uparrow_{\text{z, QND2}}\vert\uparrow_{\text{z, QND1}}\land\text{click})=90.7\%$ is observed.

To verify the preparation of single photons with our two QND detectors from the initial coherent pulse, we employ two absorbing detectors in a Hanbury Brown-Twiss configuration as shown in Fig. \ref{fig:setup}(b). This allows us to extract the second-order photon-correlation function $g^{(2)}(\tau)$. For the measurement, we use an initial coherent pulse with an average photon number of $\braket{n}=0.45$. We extract $g^{(2)}(0)$ and $\overline{g}^{(2)}(\tau\ne 0)$ averaged over all $\tau\ne0$. The second-order correlation function is conditioned on clicks in the two QND measurements. Table \ref{tab:g2data} contains the obtained data.
\begin{table}[t]
\caption{\label{tab:g2data} Measurements of the second-order photon-correlation function $g^{(2)}(\tau)$. The obtained data is conditioned on different combinations of QND detection events. Error bars are statistical.}
\begin{tabular}{ c c c } 
 \hline\hline
 Condition & $g^{(2)}(\tau=0)$ & $\overline{g}^{(2)}(\tau\ne 0)$ \\
 \hline
 \text{None} & $1.005^{+0.253}_{-0.206}$ & $0.995^{+0.006}_{-0.005}$ \\ 
 $\uparrow_{z,\text{QND} 1}$ & $0.354^{+0.121}_{-0.091}$ & $1.010^{+0.004}_{-0.004}$ \\ 
 $\uparrow_{z,\text{QND} 2}$ & $0.047^{+0.021}_{-0.015}$ & $1.000^{+0.002}_{-0.002}$ \\
 $\uparrow_{z,\text{QND} 1} \land \uparrow_{z,\text{QND} 2}$ & $0.038^{+0.023}_{-0.014}$ & $1.000^{+0.002}_{-0.002}$\\
 \hline\hline
\end{tabular}
\end{table}
When not conditioning on any of the QND detection results, we verify the coherent character of our light since $g^{(2)}(\tau)$ is close to unity for all values of $\tau$. As a next step, we condition our data on a click of QND1 and observe a reduction of $g^{(2)}(0)$ from unity to $0.354^{+0.121}_{-0.091}$. Similarly, when conditioning the data on a click of QND2, we obtain $g^{(2)}(0)=0.047^{+0.021}_{-0.015}$. Both results show the single-photon character of the light after conditioning on the respective detector. The difference in the two obtained values for $g^{(2)}(0)$ stems from the different mean photon number in the coherent pulse impinging onto the respective QND detector due to the losses between the detectors. Finally, when the data is conditioned on a click of both QND detectors, we obtain $g^{(2)}(0)=0.038^{+0.023}_{-0.014}$. A comparison between the conditions $\uparrow_{\text{z,QND1}}$ and $\uparrow_{\text{z,QND1}}\land\uparrow_{\text{z,QND2}}$ shows that the single-photon character of the light after successful state preparation with QND1 is preserved by QND2. 

In conclusion, we have shown the feasibility of repeatedly detecting an optical photon. This feature of a QND measurement allows us to enhance the detection efficiency or the signal-to-noise ratio. By adding time resolution to the QND detector, it should be possible to gain information about the propagation direction of a photon, a piece of information only accessible with QND detectors. Moreover, an improved setup with smaller losses between the two detectors could serve as a heralded source of photonic Fock states. Such a \textit{gedanken} device, described in detail in the \hyperref[supplement]{Supplemental Material}, is capable of decomposing a coherent state into its number-state constituents \cite{guerlin2007}. Arguably most fascinating is an extension to several nondestructive detectors for photonic qubits. This could speed up a plethora of quantum-network \cite{kimble2008,reiserer2015,wehner2018} protocols such as entanglement distribution where a new transmission attempt could be started immediately after a spatially-resolved loss detection between sender and receiver \cite{niemietz2021}.
\begin{acknowledgments}
This work was supported by the Bundesministerium f\"{u}r Bildung und Forschung via the Verbund Q.Link.X (16KIS0870), by the Deutsche Forschungsgemeinschaft under Germany’s Excellence Strategy – EXC-2111 – 390814868, and by the European Union’s Horizon 2020 research and innovation programme via the project Quantum Internet Alliance (QIA, GA No. 820445). E.D. acknowledges support by the Cellex-ICFO-MPQ postdoctoral fellowship program in the early stages of the experiment.
\end{acknowledgments}

\clearpage

\section{SUPPLEMENTAL MATERIAL}
\label{supplement}

\section{Theoretical modelling with the QuTip toolkit}
\noindent We model the experimental results with the QuTip toolkit for Python \cite{johansson2012}. The model is based on cavity input-output theory as outlined in \cite{kuhn2015}. Generally, the light reflected from each QND detector can be scattered into four different output modes, namely the reflection or transmission through the cavity, scattering via the atom and scattering on the cavity mirrors. We model the entire system by a concatenation of beam splitters describing the scattering into the respective modes. To account for the phase shift of the reflected light, the corresponding beam splitter is implemented using an additional phase shifter. Also, a lossy depolarizing quantum channel between the two setups is taken into account in our simulation. After propagating the light through the entire system of beam splitters and phase shifters, we trace out the loss channels and just consider the mode reflected from both cavities. Eventually, the light is detected with absorbing single-photon detectors which are modelled with another beam splitter to take into account the limited detection efficiency. Dark counts in the detectors are modelled by supplying one port of this beam splitter with thermal noise. In our simulation, we propagate the atom-atom-photon state through the series of beam splitters and eventually obtain a final density matrix.

Parameters extracted in independent characterization measurements are used in the theoretical model. These parameters comprise the fidelity of state preparation and readout of the atoms ($99\%$ at each of the nodes), their coherence times ($420\,\mathrm{\mu s}$ and $470\,\mathrm{\mu s}$, respectively), the transmission losses through the entire system ($60\%$ and $55\%$ intensity reflection of the two cavities, $53\%$ transmission between the setups and $50\%$ detection efficiency after QND2), the background-count rate of the absorbing detectors ($40\,\mathrm{Hz}$) and the relevant cavity QED parameters ($g_1(g_2)=7.6(7.6)\,\mathrm{MHz},\kappa_1(\kappa_2)=2.5(2.8)\,\mathrm{MHz},\gamma_1(\gamma_2)=3.0(3.0)\,\mathrm{MHz}$). Here, $g_{1}(g_{2})$ denotes the respective atom-cavity coupling constant (half the vacuum Rabi frequency) while $\kappa_{1}(\kappa_{2})$ and $\gamma_{1}(\gamma_{2})$ denote the respective cavity field decay rates and the atomic polarization decay rates. The given atom-cavity coupling constants are achieved on the cycling transition between the states $\ket{\uparrow_z}=\ket{5\isotope[2]{S}_{1/2}, F=2, m_F=2}$ and 
$\ket{e}=\ket{5\isotope[2]{P}_{3/2}, F=3, m_F=3}$. Both of our atom-cavity systems operate in the strong-coupling regime and we achieve a cooperativity of $C_1(C_2)=3.9(3.4)$ where $C_{1(2)}=g_{1(2)}^2/(2\kappa_{1(2)}\gamma_{1(2)})$. Besides the aforementioned experimental parameters, our theoretical model also takes the depolarization in the connection fiber as well as a residual fiber birefringence into account. 
\section{Information on measurement time, total number of experimental runs, and coincidence-click rates}
Depending on the employed mean photon number $\braket{n}$, we use a different total measurement time to obtain an error bar on the order of $1\%$ on the respective measured click probabilities of our two QND detectors. For the lowest employed photon number of $\braket{n}=0.04$, we measured for $199.8\,\mathrm{min}$ while for the highest $\braket{n}=3.11$, we only measured for $13.4\,\mathrm{min}$. In total, the measurement time for the entire scan of $\braket{n}$ amounts to $6.7$ hours. In some cases, the obtained data is conditioned on the detection event on the classical HBT photon detectors. Depending on the mean photon number employed, we achieve click rates between $0.12\,\mathrm{Hz}$ for $\braket{n}=0.04$ and $8.09\,\mathrm{Hz}$ for $\braket{n}=3.11$ on these classical detectors. These click rates are raw rates and not conditioned on the presence of a well pumped atom in each of the cavities. We start an experimental run of our double QND detection only if one atom is available in each of the cavities. The total number of experimental runs we perform until the error bars on the measured probabilities are on the order of $1\%$ depends on the mean photon number employed. In our experimental realization, the total number of runs varied between $4.7\times10^5$ for $\braket{n}=0.04$ and $3.0\times10^4$ for $\braket{n}=3.11$. Of particular importance in the experiment is the rate of events in that both QND detectors click conditioned on the detection click in the HBT detector. For the lowest employed mean photon number $\braket{n}=0.04$, we observed 1068 such events in $199.8\,\mathrm{min}$ which corresponds to a rate of $0.09\,\mathrm{Hz}$. For the highest photon number of $\braket{n}=3.11$, we observed 2155 such events in $13.4\,\mathrm{min}$ which gives a coincidence-click rate of $2.68\,\mathrm{Hz}$.

\section{Individual characterization of the two QND detectors}
\noindent We characterize our two QND devices individually. For the measurements, both atoms are initialized in the state $\ket{\uparrow_z}$ via optical pumping for $50\,\mathrm{\mu s}$ with right-circular polarized light resonant with the $\ket{\uparrow_z}\leftrightarrow \ket{e}$ transition. This pumping light is impinging on the cavity axis and transmitted through the cavity. After the pumping, the atoms are prepared in $\ket{\uparrow_x}$ by applying a $4\,\mathrm{\mu s}$ long Raman pulse with an area of $\pi/2$. Subsequently, a coherent pulse $\ket{\alpha}$ is reflected successively from the two atom-cavity systems and then impinges on absorbing photon detectors. The employed absorbing detectors are superconducting nanowire devices with a quantum efficiency of $\eta=90\%$ at our wavelength of $\lambda=780\,\mathrm{nm}$. They are connected to the output of the circulator at QND2 with a 40-m-long optical fiber. The optical pulse $\ket{\alpha}$ is propagating in a $60\,\mathrm{m}$ single mode fiber between the two resonator systems. The spatial extent of the pulse in the fiber is $200\,\mathrm{m}$, larger than the separation of the two QND detectors. Due to an engineered AC Stark shift of the excited states in our atoms, our scheme is capable of detecting photons in a specific polarization state, namely in a right-circular polarization \cite{hacker2019supplement}. To ensure that this particular polarization is maintained throughout the fiber, the birefringence of the latter is actively stabilized with piezoelectric fiber squeezers \cite{rosenfeld2008supplement} such that a one-to-one polarization mapping between the two cavities is established. After reflecting the light from the two cavities, we apply a $\pi/2$ pulse to the atoms such that they end up in the state $\ket{\uparrow_z}$ if an an odd number of photons impinges. In case of an even photon number, the atoms end up in the state $\ket{\downarrow_z}$. Atomic state detection eventually allows us to distinguish between $\ket{\uparrow_z}$ and $\ket{\downarrow_z}$ with a high fidelity ($>99\%$). After the atomic readout, we apply cooling light to the atoms for $660\,\mathrm{\mu s}$ and repeat the entire protocol with a rate of $1\,\mathrm{kHz}$.

For vanishing $\braket{n}$, an atom in the state $\ket{\uparrow_z}$ heralds the presence of a single photon. In the range of low values of $\braket{n}$, we achieve maximal detection efficiencies of $\text{P}(\uparrow_{\text{z,QND1}}\vert\text{click})=81.3\%$ and $\text{P}(\uparrow_{\text{z,QND2}}\vert\text{click})=87.0\%$ with our two QND detectors. With increasing $\braket{n}$, the probabilities $\text{P}(\uparrow_{\text{z,QND1(2)}}\vert\text{click})$ decrease, due to higher photon-number contributions in the coherent pulse, and eventually saturate at a level of $50\%$. The saturation behavior stems from the weight of even and odd photon-number contributions in the coherent pulse becoming equal when $\braket{n}$ is increased.
\renewcommand{\thefigure}{S1}
\begin{figure}[b]
\centering
\includegraphics[width=\columnwidth]{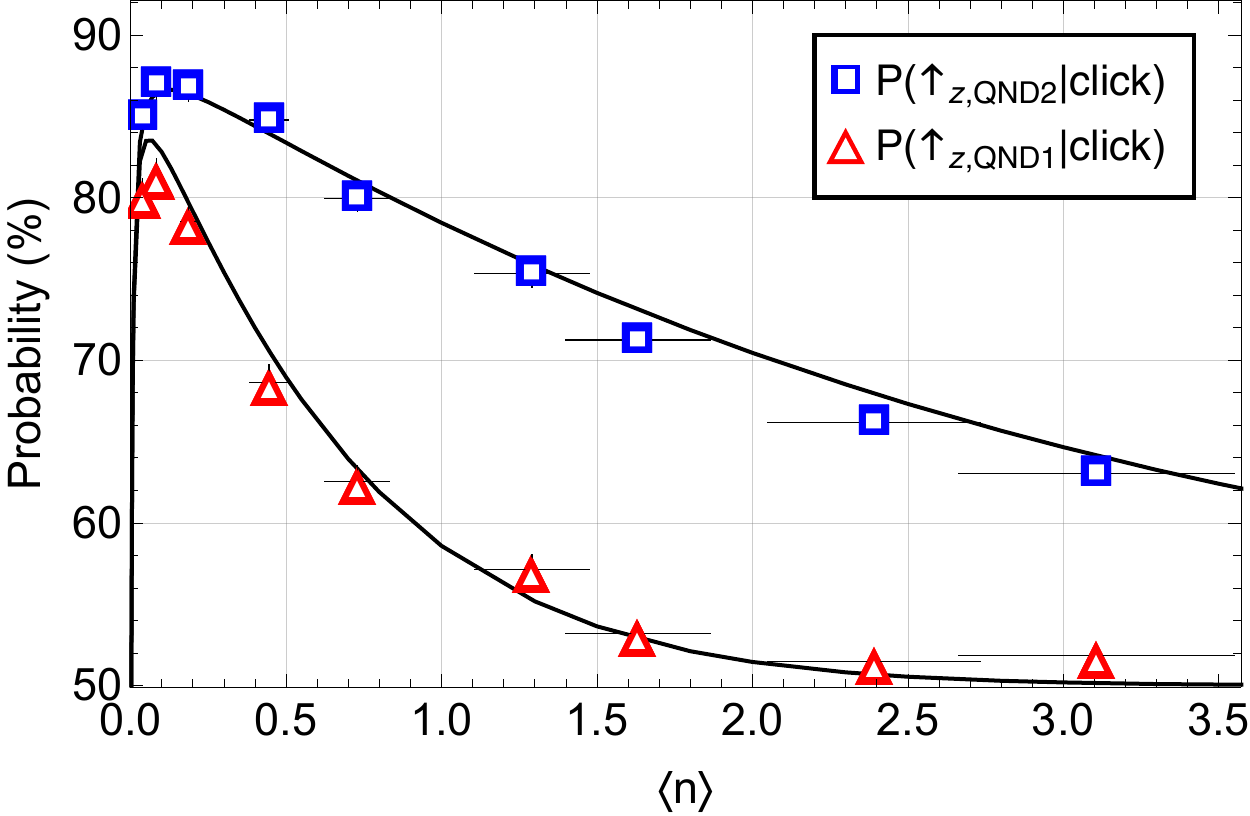} 
\caption{\label{fig:data1} Individual characterization of our two QND detectors. We show the probabilities $P(\uparrow_{\text{z,QND1/2}}\vert\text{click})$ as red triangles/blue squares. The value for $\braket{n}$ on the horizontal axis describes the mean photon-number in the pulse impinging onto the first QND detector. The solid lines are the result of our theoretical model. The error bars represent standard deviations from the mean.}
\end{figure}
Since considerable losses occur between the two systems ($53\%$ transmission between the two QND detectors), the mean number of photons impinging on the second cavity is considerably lower than on the first cavity. The data characterizing the two QND detectors is shown in Fig. \ref{fig:data1}. Due to the photon losses between the setups, we observe a different scaling behavior of the probability $\text{P}(\uparrow_z\vert \text{click})$ for the two different detectors.

\section{Proposal: A heralded source of photon-number states}
\noindent As an outlook, we describe how our setup could be employed as a heralded source of photonic number states (Fock states) by decomposing incoming coherent states. The described scheme works up to photon numbers of $n=3$ with two atom-cavity systems, but can be extended to higher photon numbers with additional atom-cavity systems. In general, coherent states containing up to $2^{k}-1$ photons can be decomposed into Fock states with $k$ concatenated systems. In the following discussion, we concentrate on the case $k=2$. 

In our experiment, a weak coherent pulse $\ket{\alpha}$ is reflected successively from the two cavities. We assume that the pulses are weak and that photon-number contributions with $n>3$ are negligible. Thus, the input state can be written as $\alpha\ket{0}+\beta\ket{1}+\gamma\ket{2}+\delta\ket{3}$ with $\vert\alpha\vert^2+\vert\beta\vert^2+\vert\gamma\vert^2+\vert\delta\vert^2\approx 1$. After optical pumping of the atoms in QND1 and QND2 into the state $\ket{\uparrow_z}$, a qubit rotation around the y axis is applied that brings them into the state $\ket{\uparrow_x}$. This qubit rotation can be expressed in matrix form as \cite{nielsen2002} 
\begin{equation}
\text{R}_y(\theta)=\begin{pmatrix}\cos\theta/2 & -\sin\theta/2 \\\sin\theta/2 & \cos\theta/2\\\end{pmatrix}\overset{\theta=\pi/2}{=}\frac{1}{\sqrt{2}}\begin{pmatrix}1 & -1 \\1 & 1\\\end{pmatrix}.
\end{equation} The cavity of the first node is tuned into resonance with the $\ket{\uparrow_z}\leftrightarrow \ket{e}$ transition such that even/odd photon numbers result in a phase shift of even/odd multiples of $\pi$, toggling the atomic superposition into $\ket{\uparrow_x}$ or $\ket{\downarrow_x}$.
\renewcommand{\thefigure}{S2}
\begin{figure}[t]
\centering
\includegraphics[width=\columnwidth]{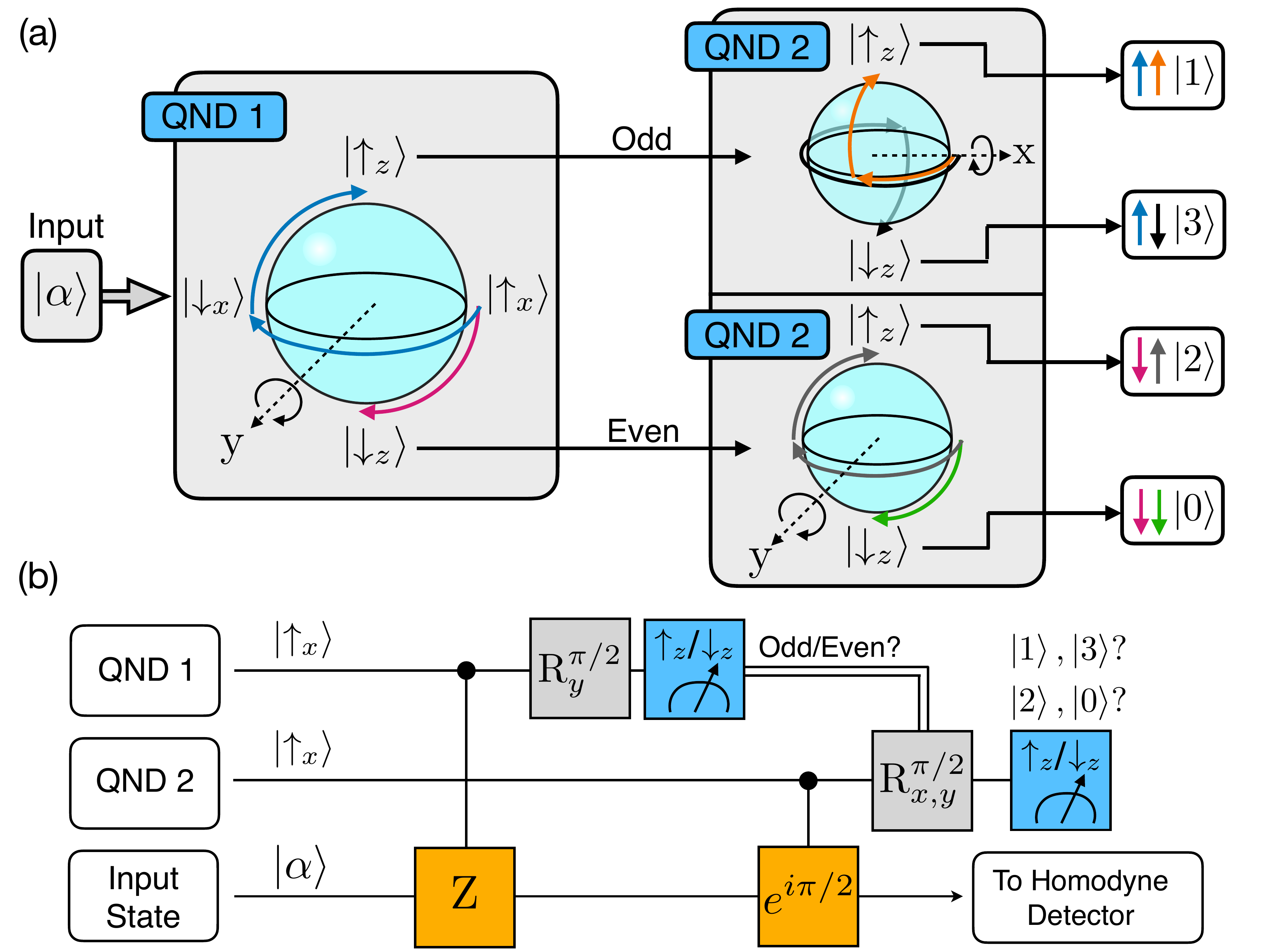}
\caption{\label{fig:photonsorter} (a) Dichotomy tree of the heralded source of photon-number states with both atoms initially prepared in the state $\ket{\uparrow_x}$. Toggling of the atomic superposition state through the photons rotates the atomic state on the equator of the Bloch sphere. QND1 allows to distinguish between even and odd photon numbers by transferring phase information to a population difference with a $\pi/2$ pulse around the y-axis. Together with the information from the first QND detector, QND2 can then fully distinguish between the four possible photon numbers by applying a $\pi/2$ pulse around the x- or y-axis, depending on the measurement outcome in QND1. (b) Circuit diagram for a heralded source of Fock states. After initialization of the atoms, the light is successively reflected from the two QND detectors. The state-detection outcome of the first atomic qubit is used as a classical signal to set the rotation axis for the state-detection basis preparation of the second atom. Eventually, the light can be analyzed with a homodyne detection setup \cite{hacker2019supplement} or alternatively be sent to a receiver for further use.}
\end{figure}
After the reflection of the light from the first node, the atom in this node is subject to another $\text{R}_\text{y}(\pi/2)$ rotation which maps $\ket{\uparrow_x}$ $(\ket{\downarrow_x})$ to $\ket{\downarrow_z}$ $(\ket{\uparrow_z})$. A final state detection of the atom thus heralds an even (odd) photon number, if the atom is found in $\ket{\downarrow_z}$ $(\ket{\uparrow_z})$ \cite{hacker2019supplement}.

The purpose of the second cavity is to discriminate between the two possible even ($n=0$ or $2$) and odd ($n=1$ or $3$) photon numbers. For that, the second cavity is detuned \cite{hacker2019supplement} such that a phase shift of $\pi/2(\pi)$ results for a single photon (two photons). The light coming from QND1 is reflected from this cavity and after the reflection, a qubit rotation pulse is applied. This pulse has an area of $\pi/2$ and its rotation axis depends on the measurement outcome of the state detection on the first atom. If an even photon number is detected at the first detector, a $\text{R}_\text{y}(\pi/2)$ rotation is applied to the second atom and maps $\ket{\uparrow_x}$ $(\ket{\downarrow_x})$ to $\ket{\downarrow_z}$ $(\ket{\uparrow_z})$. A final state detection of this atom in $\ket{\downarrow_z}$ $(\ket{\uparrow_z})$ therefore heralds the photonic Fock states $\ket{0}$ $(\ket{2})$.
If an odd photon number is detected at the first node, a $\pi/2$ qubit rotation around the x axis is applied to the second atom. It can be expressed as 
\begin{equation}
\text{R}_x(\theta)=
\begin{pmatrix}\cos\theta/2 & -i\sin\theta/2 \\-i\sin\theta/2 & \cos\theta/2\\\end{pmatrix}\overset{\theta=\pi/2}{=}\frac{1}{\sqrt{2}}\begin{pmatrix}1 & -i \\-i & 1\\
\end{pmatrix}.
\end{equation}
This pulse maps $\ket{\uparrow_y}=\frac{1}{\sqrt{2}}(\ket{\uparrow_z}+i\ket{\downarrow_z})$ to $\ket{\uparrow_z}$ and $\ket{\downarrow_y}=\frac{1}{\sqrt{2}}(\ket{\uparrow_z}-i\ket{\downarrow_z})$ to $\ket{\downarrow_z}$. A final state detection of this atom in $\ket{\uparrow_z}$ $(\ket{\downarrow_z})$ therefore heralds the Fock states $\ket{1}$ $(\ket{3})$. The final light state can be characterized by employing a homodyne detection to reconstruct its Wigner function \cite{hacker2019supplement}.
Fig. \ref{fig:photonsorter} shows the dichotomy tree and the quantum circuit diagram of the devised protocol. It should be noted that the losses between the two setups need to be lowered substantially for this protocol. Furthermore, improved cavities with higher reflectivities are necessary. Nevertheless, the protocol is appealing since it can provide an elegant tool to decompose a coherent state into its different Fock-state contributions in a heralded fashion.

\end{document}